\documentclass[prb,aps,twocolumn,superscriptaddress]{revtex4}
\usepackage{graphicx,bm,amsmath,amssymb,psfrag,color}
\usepackage{multirow}
\usepackage{xcolor}
\usepackage[normalem]{ulem}  

\begin{document}
\title{Anisotropic transport properties in prismatic topological insulator nanowires}
\author{Hallmann Ó.\ Gestsson}
\affiliation{Dept.\ of Engineering, Reykjavik University, Menntavegi 1, IS-102 Reykjavik, Iceland}
\affiliation{Dept.\ of Physics and Astronomy, University College London, Gower Street, London WC1E 6BT}
\author{Andrei Manolescu}
\affiliation{Dept.\ of Engineering, Reykjavik University, Menntavegi 1, IS-102 Reykjavik, Iceland}
\author{Jens H. Bardarson}
\affiliation{Dept.\ of Physics, KTH Royal Institute of Technology, Stockholm, SE-106 91 Sweden}
\author{Sigurdur I.\ Erlingsson}
\affiliation{Dept.\ of Engineering, Reykjavik University, Menntavegi 1, IS-102 Reykjavik, Iceland}

\begin{abstract}
The surface of a three dimensional topological insulator (TI) hosts surface states whose properties are determined by a Dirac-like equation.
The electronic system on the surface of TI nanowires with polygonal cross-sectional shape 
adopts the corresponding polygonal shape.
In a {\em constant} transverse magnetic field, such an electronic system exhibits rich properties as different facets of the polygon experience different values of the magnetic field due to the changing magnetic field projection between facets.
We investigate the energy spectrum and transport properties of nanowires where we consider three different polygonal shapes, all showing distinct properties visible in the energy spectrum {\em and} transport properties.  
Here we propose that the wire conductance can be used to differentiate between cross-sectional {\em shapes} of the nanowire by rotating the magnetic field around the wire.  Distinguishing between the different shapes also works in the presence of impurities as long as conductance steps are discernible, thus revealing the sub-band structure. 
\end{abstract}
\maketitle

\section{Introduction}
Properties of charged particles moving in two dimensions (2D) in a constant magnetic field are described by Landau levels, and this holds both for particles governed by the Schrödinger equation and the Dirac equation.  
The corresponding energy spectra do not depend on the {\em direction} of the magnetic field, only on the magnitude of the perpendicular projection of the field, in the absence of Zeeman coupling.
In systems where the magnetic field projection normal to the surface changes sign, interesting physics can occur, e.g., the presence of states with trajectories along the spatial lines of zero normal magnetic field. Such so-called snake states have been studied in nanostructures, first for semiconductors in the integer quantum Hall regime \cite{mueller92:385,reijniers00:9771,Zwerschke1999}, and later in graphene \cite{oroszlany08:081403}. 

Snake states can also occur in a {\em constant} magnetic field if the relative orientation of the electronic system changes with respect to the direction of the field, which is the situation for electrons localized on the surface of a nanowire \cite{Tserkovnyak2005,Ferrari2008,Manolescu2013,Rosdahl2015,manolescu16:205445,Chang2017}.  
For semiconducting nanowires the electronic system may be engineered to follow the surface using Fermi level pinning \cite{Heedt2016}, or a core-shell heterostructure with an insulating core and a conducting shell \cite{rieger12:5559,fan06:5157,heurlin15:2462}.  
\begin{figure}[t]
\centering
\includegraphics[width=0.48\textwidth]{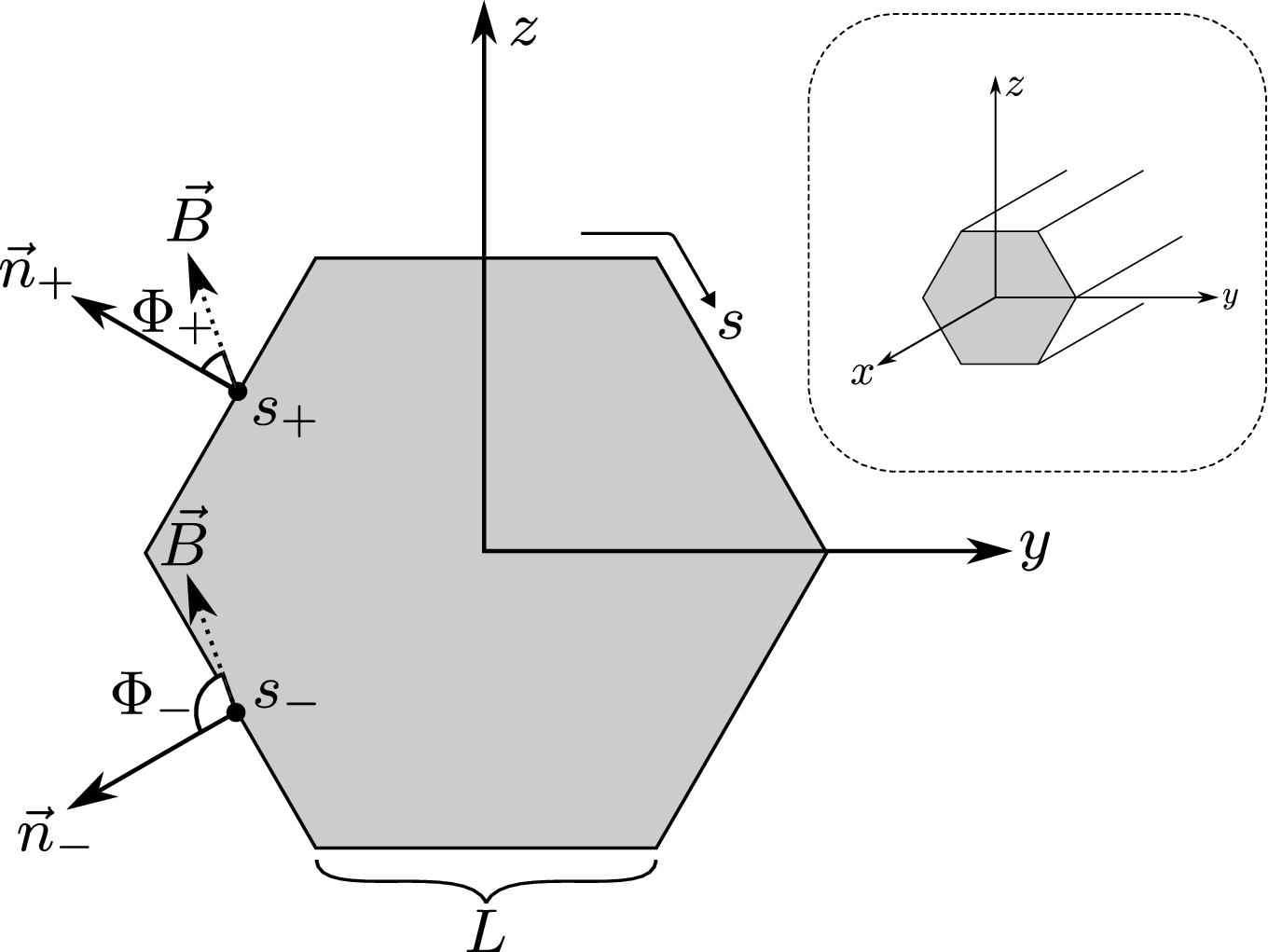}
\caption{A nanowire assuming a hexagonal cross section. We take the $x$- and $s$-coordinates to parameterize the longitudinal direction of the wire and its perimeter, respectively. The vectors $\vec{B}$ and $\vec{n}$ represent the magnetic field and surface normal, respectively, and $\Phi$ is the angle between them.  Points $s_\pm$ are midpoints on corresponding facets with fixed normal $\vec{n}_\pm$.  Facet length $L$ is fixed.
}
 \label{fig:Fig1}
\end{figure}
An interesting feature of the semiconductor nanowires is that the cross section typically assumes a polygonal shape, e.g., hexagonal.
Notably, if the electrons are constrained to such a narrow shell, the lowest energy states become localized at the corners of the polygonal cross section for Schrödinger-like systems \cite{Ballester2012,Royo2014,Sitek2015},
while in a magnetic field
the electronic states show properties that can change from Landau levels to snaking states depending on which facet the electrons are located \cite{Ferrari2009}.
Such nanowires can be considered tubular conductors, with a prismatic geometry, and their electrical conductance in the presence of a transverse magnetic field is expected to depend on the relative angles between the magnetic field and the facets of the nanowire surface \cite{torres18:085419}.  
However, to the best of our knowledge, this kind of anisotropic conductivity has not been reported in experimental studies.

The previously mentioned examples referred to systems described by the Schr\"odinger equation. In more recent years, topological insulator (TI) \cite{kane05:226801,hasan10:3045,qi11:1057,ando13:102001} have come to the forefront as materials that host surface states in nanowires \cite{dufouleur13:186806,cho15:7634,jauregui16:345,dufouleur17:45276,Munning2021Feb,roessler23:2846}.
This is due to the fact that TIs are characterized by an insulating bulk (either 3D or 2D), but conducting surface states (2D or 1D, respectively) whose low energy properties are described by the Dirac Hamiltonian, with a linear energy-momentum relation  \cite{qi11:1057,liu10:045122}. 
These nanowires show rich Aharanov-Bohm related phenomena \cite{bardarson10:156803,zhang10:206601,rosenberg10:041104,Bardarson13:056501,dufouleur18:075401} for magnetic fields parallel to the nanowire axis  \cite{dufouleur13:186806,cho15:7634,jauregui16:345,dufouleur17:45276,Munning2021Feb,roessler23:2846}, or Landau level and snake-like states for transverse fields  \cite{zhang11:015004,Shevtsov12:031004,tang14:964,vafek11:245417,brey14:085305,ilan2015:096802,xypakis17:035415,erlingsson17:036804} that can give rise to rich transport phenomena, e.g., reversal of thermoelectric currents \cite{erlingsson17:036804}.
Effective 2D models for TI nanowires have been used to investigate, e.g., Andreev reflection in T-junctions \cite{Fuchs2021Aug}, or cone shaped nanowires \cite{PhysRevLett.124.126804}.  All the 2D models assume a fixed Fermi velocity, but for example in topological crystalline insulators different terminations of the crystalline lattice can host different velocities leading to hinge and corner states \cite{nguyen22:075310,skiff23:011}.

In this paper we consider the electronic properties of TI nanowires of polygonal shape in a constant transverse magnetic field.  
Different projections of the magnetic field along different polygon facets lead to energy spectra showing unique characteristics for each structure, e.g., two separate sets of Landau levels on different facets.  
We propose that the conductance of the wire, as a function of magnetic field orientation, 
can be used to differentiate between cross-sectional shapes of the nanowires.
To model the transport we introduce a modified Wilson mass term which faithfully reproduces the correct Landau levels on different facets.  We use the recursive Green's function method to calculate the transmission function and show that for low enough impurity densities the anisotropic conductance reveals the underlying nanowire shape.

The structure of the paper is as follows: Following the Introduction, the model system and Hamiltonian is discussed in Sec.\ \ref{sec:model}.  We also outline the Landau level approximation for general facets of the polygon in Sec.\ \ref{sec:LLA}.  The modification to the Wilson mass term for transport calculation is outlined in Sec.\ \ref{sec:wilson} along with transmission function calculations.  The conductance anisotropy is discussed in Sec. \ref{sec:cond}, and finally we discuss the results in Sec.\ \ref{sec:conc}.

\section{Model Hamiltonian and facet Landau levels}
\label{sec:model}
We consider a nanowire composed of a TI material such that the wire surface can support edge modes, i.e., quasi-particles which propagate along the wire surface whilst the bulk remains insulating. 
Furthermore, we assume that there is a constant magnetic field perpendicular to the length of the wire. 
The wire surface is taken to be parameterized by $x$ and $s$, where $x$ spans the entire length of the wire and $s$ is the arc-length variable for the cross section perimeter of the wire, as shown in Fig.\ \ref{fig:Fig1}. 
The effective Hamiltonian describing the surface modes of such a system is \cite{PhysRevLett.105.036803,PhysRevLett.105.206601,deJuan19:60}
\begin{equation}
H_\mathrm{surf}(x,s) = v_F\left \{ \sigma_x\left [p_x + eA_x(s)\right ] +\sigma_y p_s\right \},
\label{eq:Hsurf}
\end{equation}
where $p_s=-i \hbar \partial_s$ is the momentum operator of the perimeter variable $s$. 
Here $v_F$ is the Fermi velocity, $e$ is the electronic charge, $\sigma_x$ and $\sigma_y$ are the Pauli matrices, $p_x$ is the  momentum operator along the length of the wire, and $A_x$ is the $x$-component of the vector field. 
We choose the gauge such that $A_y=A_z=0$, and
\begin{equation}
A_x(y,z)=B[ \sin ( \phi) z  -\cos (\phi) y],
\label{eq:Ax}
\end{equation}
where $B = \vert\vec{B}\vert$ and $\phi$ is the angle between $\vec{B}$ and the $z$-axis. 
For the linear-in-momentum surface Hamiltonian
the polygon corners have no effect on the surface states \cite{lee09:196804,vafek11:245417,brey14:085305,messiasDeResende17:161113} and the influence of the polygonal shape only enters via discontinuities in the slope of $A_x(s)$.

In the absence of a magnetic field $H_\mathrm{surf}$ admits a simple analytical spectrum. Squaring the Hamiltonian in Eq.\ (\ref{eq:Hsurf}) and using $\{\sigma_x,\sigma_y\}=0$ leads to 
\begin{equation}
H_\mathrm{surf,0}^2 = \hbar^2v_F^2\left(p_x^2 + p_s^2\right)\sigma_0,
\label{eq:Hsurf2}
\end{equation}
with $\sigma_0$ being the $2\times 2$ identity matrix.  Applying the plane wave ansatz $\psi(x,s) = \frac{1}{\sqrt{LP}}e^{i(k_xx + k_ss)}$, where $L$ is the length of the wire and $P$ is the cross section perimeter length, yields eigenvalues of the form $E_\pm = \pm\hbar v_F\sqrt{k_x^2 + k_s^2}$. 
A nontrivial spin connection has been gauged away from the Hamiltonian and into the boundary condition
$\psi(x,s+P)=-\psi(x,s)$, leading to $k_s=\frac{2\pi}{P}\left(n+\frac{1}{2}\right)$ with integer $n$ \cite{bardarson10:156803,zhang10:206601,rosenberg10:041104,bardarson18:93}. 

\subsection{Facet Landau levels}
\label{sec:LLA}
In the presence of a perpendicular magnetic field the vector potential $A_x(s)$ changes according to the form of the prismatic nanowire.  
For translationally invariant wires where $[p_x,H_\mathrm{surf}]=0$, $p_x$ can be replaced by its eigenvalue $\hbar k_x$ in Eq.\ \eqref{eq:Hsurf}. 
\begin{figure}[tb]
\centering
\includegraphics[width=0.48\textwidth]{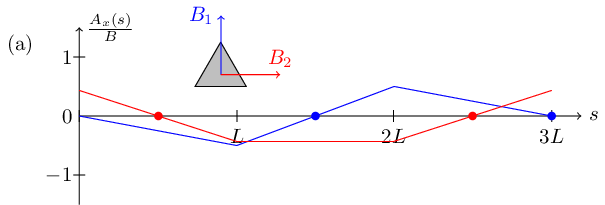}
\includegraphics[width=0.48\textwidth]{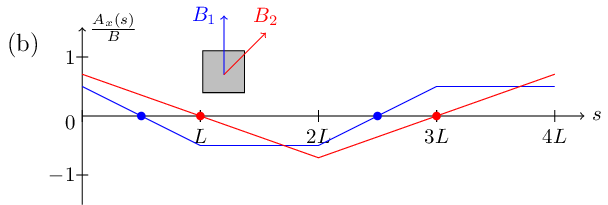}
\includegraphics[width=0.48\textwidth]{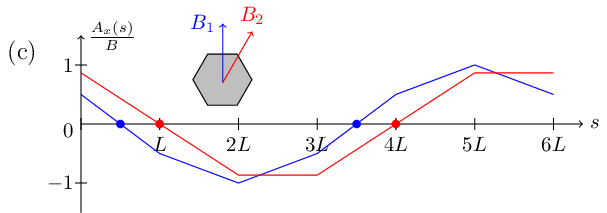}
\caption{Normalized vector potential, $A_x(s)/LB$, as a function of the perimeter variable $s$ for two different orientations of the magnetic field for different cross sections and orientation: 
a) triangle for $\phi=0$, and $\phi=\pi/2$, b) square $\phi=0$, $\phi=\pi/4$, and c) hexagon for $\phi=0$ and $\phi=\pi/6$.}
\label{fig:Fig2}
\end{figure}
Depending on the orientation of the magnetic field, and the local normal $\vec{n}(s)$, Landau levels can form at different facets, at different values of $k_x$.  
To illustrate how this occurs we consider the hexagonal nanowire with a magnetic field tilted from the $z$-axis by an angle $\phi$, and $\Phi$ is the angle between $\vec{B}$ and $\vec{n}$. 
Focusing on the tilted facets characterized by
the two angles $\Phi_+$ and $\Phi_-$, see Fig.\ \ref{fig:Fig1}, the vector potential can be linearized around the center on each facet, indicated by $s_\pm$, resulting in  
\begin{eqnarray}
A^\pm_x(s)&=&-B\cos(\Phi_\pm)(s-s_\pm)+A_x(s_\pm). \label{eq:Axp} 
\end{eqnarray}
The resulting linearized version of Eq.\ (\ref{eq:Hsurf}) is then
\begin{eqnarray}
H^{[\pm]}_\mathrm{surf}&=&v_F\left \{ \sigma_x\left [- eB^\pm
(s-s_\pm)+\hbar k_x +eA_x(s_\pm)
\right ] \right . \nonumber  \\
& &\left . \quad +\sigma_y p_s\right \},
 \label{eq:HsurfLinear}
\end{eqnarray}
where $B^\pm= B{\cos(\Phi_\pm)}$.  Landau levels centered on $s_\pm$ will form at $k_x$-values that satisfy $\hbar k_x +A_x(s_\pm)=0$, see App.\ \ref{app:Ladder}.  Moving away from these $k_x$-values by $\Delta k_x$ shifts the Landau level center coordinate by 
$\Delta s_\pm =\frac{\hbar }{e B^\pm} \Delta k_x$.  

The properties of $A_x$ as a function of the perimeter variable $s$ are shown in Fig.\ \ref{fig:Fig2}, for a) a triangle, b) a square, and c) a hexagon. 
The coordinate system is chosen such that the $x$-axis (the longitudinal direction) intersects the centroid of the nanowire, and $s=0$ is placed at the top left corner of each shape, and increases in the clockwise direction.  The slope of each piece-wise linear part gives the normal projection of the magnetic field on the corresponding facet.
Two different orientations of magnetic fields are shown for each shape. The blue and red dots indicate the position of the $s$-points (for $k_x=0$), where Landau levels will be centered, which depends on the cross section shape and magnetic field orientation.
Focusing on one such $k_x$ value, e.g., for $s_+$ the Hamiltonian can be written as
\begin{equation}
H^{[+]}_\mathrm{surf}=\frac{\hbar v_F \sqrt{|\cos ( \Phi_+)|}}{\sqrt{2}\ell_c} \left (
 a_{[+]} \sigma_+ + a_{[+]}^{\dagger} \sigma_- \right ),
\label{eq:H_plus}
\end{equation}
where $\sigma_\pm=\sigma_x \pm i\sigma_y$, $\ell_c^2=\hbar/eB$ is the magnetic length, and the ladder operator for the $\vec{n}_+$ facet is defined in in Eq.\ (\ref{eq:a_plus}) in App.\ \ref{app:Ladder}.
Being able to focus on a given facet and defining the facet Hamiltonian in Eq.\ (\ref{eq:H_plus}) relies on the wave function being nonzero only on that facet (up to exponentially small corrections).  
In terms of the magnetic length $\ell_c$ and side length $L$ this condition translates to $\ell_c /L \ll \sqrt{|\cos ( \Phi_+)|}$.
The Hamiltonian in Eq.\ (\ref{eq:H_plus}) can be diagonalized using basis states $\{  |n-1,\uparrow \rangle, |n,\downarrow \rangle \}$ for $n=1, 2, 3, \dots$ resulting in a pair of  eigenenergies (negative and positive)
\begin{equation}
\varepsilon^{[+]}_{n,\pm}= \pm \frac{\hbar v_F \sqrt{|\cos( \Phi_+)}|}{\ell_c}\sqrt{2n},
\label{eq:enp}
\end{equation}
along with a single state $|0,\downarrow \rangle$ with eigenenergy $\varepsilon^{[+]}_0=0$.
In a similar way the Hamiltonian for the $\vec{n}_-$ segment can be written as 
\begin{equation}
H^{[-]}_\mathrm{surf}=\frac{\hbar v_F \sqrt{|\cos (\Phi_-)}|}{\sqrt{2}\ell_c} \left (
a_{[-]}^{\dagger} \sigma_+  + a_{[-]} \sigma_-
\right ),
\label{eq:H_minus}
\end{equation}
where the corresponding ladder operator is defined in Eq.\ (\ref{eq:a_minus}) in App.\ \ref{app:Ladder}.
The Hamiltonian in Eq.\ (\ref{eq:H_minus}) can be diagonalized 
with the basis states
$\{  |n-1,\downarrow \rangle, |n,\uparrow \rangle \}$ for $n=1, 2, 3, \dots$, where the spin indices have been switched.  
The resulting pair of  eigenenergies (negative and positive)
\begin{equation}
\varepsilon^{[-]}_{n,\pm}= \pm \frac{\hbar v_F \sqrt{|\cos( \Phi_-)}|}{\ell_c}\sqrt{2n},
\label{eq:enm}
\end{equation}
along with a single state $|0,\uparrow \rangle$ with eigenenergy $\varepsilon^{[-]}_0=0$.
\begin{figure}[tb!]
\centering
\includegraphics[width=0.48\textwidth]{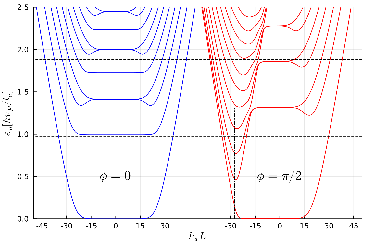}
\caption{Energy spectra for the triangle for the two magnetic field orientations shown in Fig.\ \ref{fig:Fig2}a).  The magnetic field value corresponds to $\ell_c /L=0.125$.  The two dashed lines indicate energies $E=0.97$ and $1.92$ that are relevant for subsequent transport calculations.
The vertical dash-dotted line indicates the $k_x L$-value corresponding to the  
center of the facet where the zero-magnetic field states occur, see text for details.
The zero energy Landau levels are doubly degenerate.
}
\label{fig:ET_phi}
\end{figure}

The energy spectrum, as a function of $k_x$, for a triangular nanowire is shown in Fig.\ \ref{fig:ET_phi} for the two different magnetic field orientations shown in Fig.\ \ref{fig:Fig2}a).  
Only the positive energy is shown since the negative part is simply mirrored due to an electron-hole symmetry, i.e., the anticommutation relation $ \{ H_\mathrm{surf}, \sigma_z  \} =0$ results in any state  $|\psi \rangle$ with energy $\varepsilon \geq 0$ having a corresponding state $\sigma_z |\psi \rangle$ with negative energy $-\varepsilon$.  
For  magnetic field orientation $\phi=0$ the tilted facets of the triangle have $\cos( \Phi_+)=1/2$ and on the bottom facet $|\cos( \Phi_-)|=1$.  
The spectrum shows an interesting feature of every other LL being nondegenerate at energies $\sqrt{1},\sqrt{3}, \sqrt{5}, \dots$ (in units of $\hbar v_F /\ell_c$) while LL at energies $\sqrt{0},\sqrt{2}, \sqrt{4}, \dots$  are doubly degenerate.
The LL energies, given by Eqs.\ (\ref{eq:enp}) and (\ref{eq:enm}), coincide if $|\cos( \Phi_+)| m_+=|\cos( \Phi_-)| m_-$ where $m_\pm \in \mathbb{N}$ are the LL indices for the 
$\Phi_\pm$
facets (the $\Phi_+$ facet consists of the top left and right facets).  
This happens for the even $m_+$ states on the $\Phi_+$ facet which will hybridize with the state on the opposing facet, into symmetric and anti-symmetric states, as they  evolve into linear-in $k_x$ states for $|k_x| \gg L/\ell_c^2$.
The odd $m_+$ states on the $\Phi_+$ facet {\em never} coincide with the eigenenergies of states on opposite facets.
As  $k_x$ is increased these states will be  coupled to states with different energies (2nd order perturbation) without hybridizing, as they evolve into linear-in $k_x$ states for $|k_x| \gg L/\ell_c^2$.

In the $\phi=\pi/2$ orientation the magnetic field is parallel to the bottom side, leading to a zero-magnetic field projection on that facet.  
The two tilted facets will have opposite projection and will hybridize at the apex for positive value of $k_x$.  
For negative values of $k_x$ the wave function is pushed toward the $\vec{\bm{B}}\cdot \vec{\bm{n}}=0$ facet.  This corresponds to the red curve in Fig.\ \ref{fig:Fig2}a, where one can see that $A_x(s)$  is constant for $s\in[L,2L]$. Setting $s=3L/2$, corresponding to the center of the constant region, results in a value of $k_xL$ that is determined by
\begin{equation}
\hbar k_x+e A_x(s)\bigr |_{s=3L/2}=0.
\label{eq:sZero}
\end{equation}
Note the coordinate system is placed in the centroid of the triangle which, in our chosen gauge, results in $A_x(s)=-\frac{\sqrt{3}}{4}BL$ in the constant region.
The resulting value $k_xL=\frac{\sqrt{3}L^2}{4\ell_c^2}$ is indicated by a vertical line for $\phi=\pi/2$. A set of states will form there which, to lowest order, have energies $\sim \hbar v_F\frac{\pi}{L}\times \mathrm{integer}$.  

\begin{figure}[tb]
\centering
\includegraphics[width=0.48\textwidth]{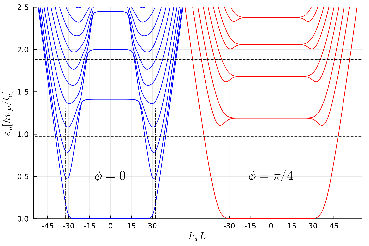}
\caption{Energy spectra for the square for the two magnetic field orientations shown in Fig.\ \ref{fig:Fig2}b).  The magnetic field value corresponds to $\ell_c /L=0.125$.  The two dashed lines indicate energies $E=0.97$ and $1.92$
that are relevant for subsequent transport calculations.
The vertical dash-dotted lines indicates the $k_x L$-values where the zero-magnetic field states occur, see text for details. The zero energy Landau levels are doubly degenerate.
}
\label{fig:ES_phi}
\end{figure}
Figure \ref{fig:ES_phi} shows the square nanowire spectra for two magnetic field orientations.  
For $\phi=0$ the two side facets have zero projection of the magnetic field so that for $|k_x L| \sim  L^2/2\ell_c^2$ the Landau levels evolve into corresponding quantized zero-magnetic field states.  
The $s$-values at the center of the flat sections of $A_x(s)$ in Fig.\ \ref{fig:Fig2}b are $s=3L/2$ and $7L/2$.  In this case, the condition in Eq.\ (\ref{eq:sZero}) results in $|k_xL|=L^2/2\ell_c^2$, indicated by the vertical dashed line.
Note that each Landau level evolves into two states on the side facet, the same applies to the zero Landau level which is doubly degenerate and splits into a positive energy state and a negative energy state (not shown).
For the $\phi=\pi/4$ orientation the magnetic field projection is never zero and takes constant value $\cos(\Phi_+)=1/\sqrt{2}$ for the  two facets with positive projection, but $\cos(\Phi_-)=-1/\sqrt{2}$ for the negative projection facets. For $|k_x L| \sim  L^2/2\ell_c^2$ the Landau levels hybridize and evolve towards a linear in $k_x$ dependence when $|k_x L| \gg L^2/\ell_c^2$.
The dispersing states occur in the two corners where $\vec{\bm{B}} \cdot \vec{\bm{n}}$ changes sign.

\begin{figure}[t]
\centering
\includegraphics[width=0.48\textwidth]{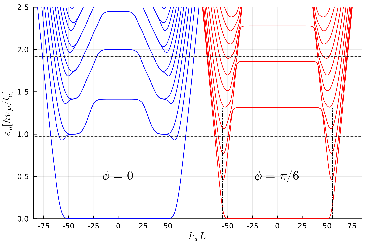}
\caption{Energy spectra for the hexagon for the two magnetic field orientations shown in Fig.\ \ref{fig:Fig2}c).  The magnetic field value corresponds to $\ell_c /L=0.125$.  The two dashed lines indicate energies $E=0.97$ and $1.92$
 that are relevant for subsequent transport calculations.
The vertical dash-dotted lines indicates the $k_x L$-values where the zero-magnetic field states occur, see text for details.
The zero energy Landau levels are doubly degenerate.
}
\label{fig:EH_phi}
\end{figure}

The hexagon is the final shape we consider.  
The energy spectra for the two orientations in Fig.\ \ref{fig:Fig2}c are shown in Fig.\ \ref{fig:EH_phi}.  
For the $\phi=\pi/6$ orientation the spectrum is similar to the $\phi=0$ spectrum for the square in Fig.\ \ref{fig:ES_phi} since on two facets $\vec{\bm{B}}\cdot \vec{\bm{n}}=0$, and on the remaining facets $\cos(\Phi_+)=|\cos(\Phi_-)|=\sqrt{3}/2$.  
The zero-magnetic field states, where $\vec{\bm{B}}$ is parallel to the facets (red curve in Fig.\ \ref{fig:Fig2}c), are determined by Eq.\ (\ref{eq:sZero}) at $s=5L/2$ and $11L/2$.  These value of $s$ on the hexagon result in $k_x L=\pm \sqrt{3}L/2 \ell_c^2$, indicated by the vertical lines.
The $\phi=0$ orientation for the hexagon shows that Landau levels can form on two facets with the same {\em sign} of $\cos(\Phi_+)$, but different values reflecting the different facet orientation.  On the top surface $\cos(\Phi_+)=1$ but on the adjacent facet $\cos(\Phi_+)=1/2$ resulting in $\varepsilon_{1,+}=1$, in units of $\frac{\hbar v_F}{\ell_c}$.
The effectively smaller magnetic field on the tilted facets translates to a smaller width of the Landau levels, i.e., smaller $k_x$ interval, before they hybridizes with the states on the facets with the opposite projection. 

The energy spectra in Figs.\ \ref{fig:ET_phi}-\ref{fig:EH_phi} are obtained by direct numerical diagonalization of Eq. (\ref{eq:Hsurf}) written in a basis of $76$ plane wave basis states, see below Eq.\ (\ref{eq:Hsurf2}), along with the two spin states.  The number of basis states is chosen such that the first 10 Landau levels (flat) conincide with the analytical energy levels for the facets given in Eqs.\ (\ref{eq:enp}) and (\ref{eq:enm}).  

\section{Lattice model in the presence of impurities}
\label{sec:wilson}
In the previous section we analyzed the Landau levels forming on different facets assuming translational invariance along the wire.  In the presence of impurities $k_x$ is no longer a good quantum number and we need to solve the Dirac eigenvalue problem in terms of both the surface variables $(x,s)$.
To model the effect of randomly distributed point impurities on the transport properties of the wire we discretize the longitudinal ($x$) variable of $H_\mathrm{surf}(x,s)$ using a finite difference approach  with a lattice constant $a$. This entails approximating the action of $\partial_x$ onto the spatial component $\psi(x,s)$ of the wave function as
\begin{equation}
\partial_x\psi(x,s) \approx \frac{1}{2a}\left [ \psi(x+a,s) - \psi(x-a,s) \right ],
\end{equation}
such that the  Hamiltonian is approximated as 
\begin{eqnarray}
    H_\mathrm{surf}\psi(x,s\!\!\!\!&\!\!)\!\!&  \!\!
  \approx  v_F\left [ e\sigma_xA_x(s) -i\hbar\sigma_y\partial_s\right ]\psi(x,s) \nonumber \\ 
  \!\!  &-&\frac{i\hbar v_F\sigma_x}{2a}\left [ \psi(x+a,s) - \psi(x-a,s)\right ].
    \label{eq:Hsurf_FD}
\end{eqnarray}
Discretizing the linear in momentum Hamiltonian in Eq.\ (\ref{eq:Hsurf}) will lead to Fermion doubling.\cite{nielsen81:219,stacey82:468} 
 For two-dimensional linearly dispersing Hamiltonians an extra {\em quadratic} term can be added to solve the Fermion doubling \cite{habib16:113105,zhou17:245137}
\begin{equation}
H_W = \frac{v_F\beta}{\hbar} \biggl \{ \sigma_z\left  [ (p_x+eA_x(s))^2 +p_s^2 \right ]
 \biggr \},
 \label{eq:H_wilson}
\end{equation}
where $\beta$ is a parameter with dimension of length. The value of $\beta=a/\sqrt{3}$ gives a linear in momentum dependence of $k_x$ over a large part of the Brillouin zone, but still lifts the Fermion doubling \cite{erlingsson18:1156}.

Although the Hamiltonian in Eq.\ (\ref{eq:H_wilson}) solves the Fermion doubling it will lead to a shift of the Landau levels, which is in opposite direction depending on the projection of $\vec{B}$ on the local normal $\vec{n}(s)$.  To illustrate this energy shift, and also show how it can remedied, we solve the facet Landau levels in the presence of $H_W$ in Eq.\ (\ref{eq:H_wilson}) for the homogeneous wire.
Starting with the definition of the ladder operators for the positive projection in Eq.\ (\ref{eq:a_plus}) and applying it to $H^{[+]}_\mathrm{surf}+H_{W}$ results in 
 \begin{eqnarray}
H^{[+]}_{\mathrm{surf},W}&=&\frac{\hbar v_F \sqrt{|\cos ( \Phi_+)|}}{\sqrt{2}\ell_c} \Biggl [
 a_{[+]} \sigma_+  +a_{[+]}^\dagger \sigma_-   \nonumber \\
+& &\!\!\!\! \!\! \sigma_z 2 \sqrt{2}\sqrt{|\cos(\Phi_+)|}\frac{\beta }{\ell_c} \left (a_{[+]}^\dagger a_{[+]} +\frac{1}{2} \right )
\Biggr ]\!,
\label{eq:HW_plus}
\end{eqnarray}
which has shifted the previously zero energy eigenstate $| 0,\downarrow \rangle$ and  the other eigenenergies are found by diagonalization in the basis$\{ |n-1,\uparrow \rangle, |n,\downarrow \rangle\}$, $n=1, 2, \dots$. 
The resulting eigenenergies are
\begin{eqnarray}
\varepsilon^{[+]}_{0}&=& -\frac{\hbar v_F }{\ell_c} \frac{\beta}{\ell_c} |\cos(\Phi_+)|  \label{eq:e0W_plus}\\
\varepsilon^{[+]}_{n,\pm}&=& \frac{\hbar v_F }{\ell_c} \left ( 
-\frac{\beta}{\ell_c} |\cos(\Phi_+)| 
 \right . \nonumber \\
&\pm& \left . \frac{\sqrt{|\cos ( \Phi_+)|}}{\sqrt{2}}
\sqrt{4n+8|\cos(\Phi_+)| \frac{\beta^2}{\ell_c^2}n^2} \right ).
\label{eq:enW_plus}
\end{eqnarray}
For the negative projection Landau levels, the same procedure starting from Eq.\ (\ref{eq:H_minus})  and the definition of ladder operators in Eq.\ (\ref{eq:a_minus}) results in the following Hamiltonian
 \begin{eqnarray}
H^{[-]}_{\mathrm{surf},W}&=&\frac{\hbar v_F \sqrt{|\cos ( \Phi_-)|}}{\sqrt{2}\ell_c} \Biggl [
a_{[-]}^\dagger \sigma_+ +a_{[-]} \sigma_-     \nonumber \\
+& &\!\!\!\! \!\! \sigma_z 2 \sqrt{2}\sqrt{|\cos(\Phi_-)|}\frac{\beta}{\ell_c} \left (a_{[-]}^\dagger a_{[-]} +\frac{1}{2} \right )
\Biggr ]\!.
\label{eq:HW_minus}
\end{eqnarray}
Now, the shifted zero energy state is $| 0,\uparrow \rangle$ and other eigenergies are obtained by diagonalizing in the basis $\{ |n-,\downarrow \rangle, |n,\uparrow \rangle\}$, $n=1,2, \dots$ 
The resulting eigenenergies are
\begin{eqnarray}
\varepsilon^{[-]}_{0}&=&\frac{\hbar v_F }{\ell_c} \frac{\beta}{\ell_c} |\cos(\Phi_-)|  \label{eq:e0W_minus}\\
\varepsilon^{[-]}_{n+1,+}&=&\frac{\hbar v_F }{\ell_c} \left ( 
\frac{\beta}{\ell_c} |\cos(\Phi_-)| 
 \right . \nonumber \\
\pm&& \!\!\!\! \left . \frac{\sqrt{|\cos ( \Phi_-)|}}{\sqrt{2}}
\sqrt{4n+8|\cos(\Phi_-)| \frac{\beta^2}{\ell_c^2}n^2} \right )\!\!.
\label{eq:enW_minus}
\end{eqnarray}
On inspection of Eqs.\ (\ref{eq:e0W_plus})-(\ref{eq:enW_plus}) and (\ref{eq:e0W_minus})-(\ref{eq:enW_minus}), one can see that the shift that breaks the Landau level degeneracy can be removed by {\em adding} a term $v_F \beta eB\cos(\Phi)=v_F \beta e \vec{B}\cdot \vec{n}(s)$. The amended Wilson mass term accounting for different orientation of surface normals with respect to the applied magnetic field is thus
\begin{eqnarray}
H_{W,B} \!&\!=\!&\! \frac{v_F\beta}{\hbar} \biggl \{   \sigma_z\left [( p_x+eA_x(s))^2  +p_s^2 \right ]
  + \sigma_0\frac{e}{\hbar}\vec{B}\cdot {\vec{n}}(s)   \biggr \} . \nonumber \\
 \label{eq:H_wilsonB}
\end{eqnarray}
\begin{figure}[tb]
\centering
\includegraphics[width=0.48\textwidth]{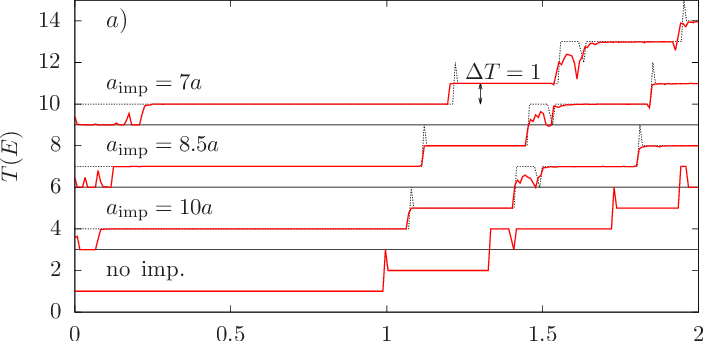} 
\includegraphics[width=0.48\textwidth]{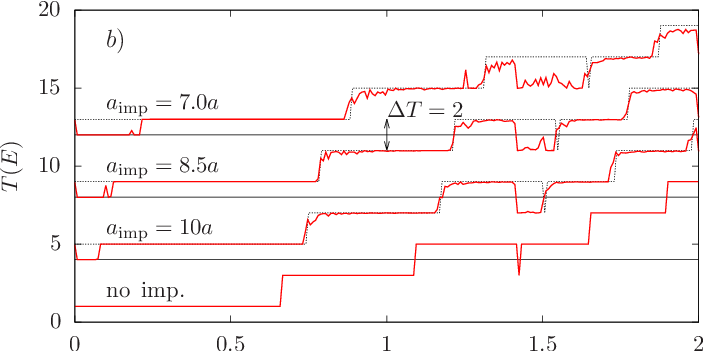}
\includegraphics[width=0.48\textwidth]{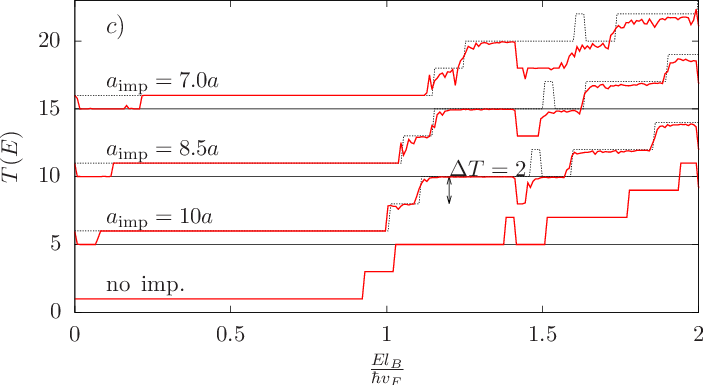}
\caption{
The transmission function $T(E)$ for a) triangle, b) square, and c) hexagon.  The curves corresponding to different impurity densities are offset for clarity.  For comparison, the gray lines show $T(E)$ for the clean wire shifted by $\Delta E=\langle V \rangle$.  The impurity strength is $V_\mathrm{imp}=0.45 \frac{\hbar v_F}{\ell_c}$, $a=0.0025L$, $N_x=2400$, and $l_c/L = 0.20$.   The arrows indicate prevalent step sizes  $\Delta T$.
}
\label{fig:T_TSH}
\end{figure}
We can now discretize $H_\mathrm{surf}+H_{W,B}$,
given by Eqs.\  (\ref{eq:Hsurf}) and (\ref{eq:H_wilsonB}), respectively, in the $x$-variable (see App.\ \ref{app:Discrete} for details) 
and model randomly placed short range impurities with
\begin{equation}
V(x,s)=\sum_{n}V_\mathrm{imp} \delta(s-s_n)\delta(x-x_n),
\end{equation}
where $V_\mathrm{imp}$ is the impurity strength.  The density of impurities is controlled by the parameter $a_\mathrm{imp}$, which gives the average distance between impurities in units of $a$.
We then use standard transport methods based on the recursive Green's functions method \cite{ferry97:book,datta95:book}.  
This is outlined in the case of circular nanowires in Ref.\  \cite{erlingsson17:036804}, which can be directly applied here by replacing the cylindrical perimeter $R\varphi$ with the equivalent polygonal perimeter variable $s$.   
From the nanowire Green's function one obtains the transmission function $T(E)$.\cite{ferry97:book}

In Fig.\ \ref{fig:T_TSH} we show the transmission function for the three cross-sectional shapes, a) triangle, b) square, and c) hexagon.  The transmission calculations are done for a single impurity configuration, i.e.\ there is no averaging over impurity configurations.
The different curves, which are offset for clarity, correspond to increased impurity concentrations, parameterized by $\frac{a_\mathrm{imp}}{a}$.
Narrow features in $T(E)$ tend to get washed out by increased impurity concentration, but for all cases considered here the quantized steps are quite evident.
Note that in Fig.\ \ref{fig:T_TSH}b (square) and  Fig.\ \ref{fig:T_TSH}c (hexagon), the transmission steps are of size $\Delta T=2$, but for the triangle in Fig.\ \ref{fig:T_TSH}a the step size is predominantly of size $\Delta T=1$. This is due to the nondegenerate Landau levels and asymmetric zero-magnetic field states in the triangle shown in Fig.\ \ref{fig:ET_phi} which contribute only one conducting channel.

Note that the curves corresponding to clean systems show no spurious levels splitting due to the Wilson mass term. Experimentally there are already transport measurements that show evidence of quantized TI nanowire states \cite{Munning2021Feb,roessler23:2846}. 
For such samples the nanowire cross-sectional shape can be identified from transport properties, as we will discuss in the next section.

\section{Conductance and magnetic field orientation}
\label{sec:cond}
\begin{figure}[t]
\centering
\includegraphics[width=0.48\textwidth]{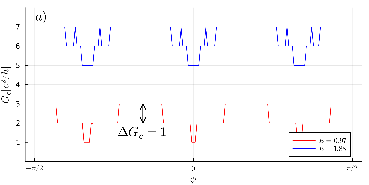}
\includegraphics[width=0.48\textwidth]{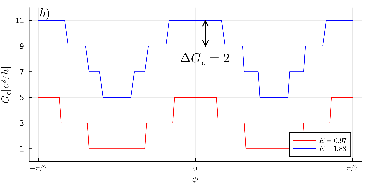}
\includegraphics[width=0.48\textwidth]{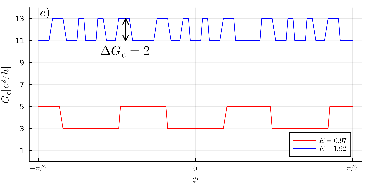}
\caption{Conductance $G_c$ as a function of angle $\phi $ for a) triangle, b) square, and c) hexagon.  The conductance is calculated at two energies for each shape, corresponding to the vertical lines in Figs.\ \ref{fig:ET_phi}-\ref{fig:EH_phi}, with $l_B/L = 0.125$.
The arrows denote the size of the conductance jumps $\Delta G_c$.
 }
\label{fig:GTSH}
\end{figure}

Having established an accurate method of calculating transmission function in TI prismatic nanowires we can make predictions regarding the two terminal conductance, and how it behaves as a function of magnetic field orientation.  In Fig.\ \ref{fig:GTSH} we plot the conductance for a) triangle, b) square, and c) hexagon, as a function of magnetic field orientation angle $\phi$.
There are two features that allow us to identify the shape of the nanowire from this: $(i)$ the number of times repeated features appear in $G_c(\phi)$ for $\phi \in [-\pi/2,\pi/2]$, and $(ii)$ how large the conductance steps are.  In Fig.\ \ref{fig:GTSH}b) the features is $G_c(\phi)$ are repeated twice, for both energies, indicating an underlying square cross-section.  
The number of repeating patterns in Figs. \ref{fig:GTSH}a) and \ref{fig:GTSH}c) is three, reflecting the underlying three-fold symmetry of both triangle and hexagon.  However, the step sizes are different: $\Delta G_c=1$ for the triangle and  $\Delta G_c=2$ for the hexagon.  The origin of the difference can be understood by comparing Fig.\ \ref{fig:ET_phi} for $\phi=\pi/2$ (triangle), and Fig.\ \ref{fig:EH_phi} for $\phi=\pi/3$ (hexagon).
For the triangle only one facet can be parallel to the applied magnetic field.  This leads to the observed asymmetry in the energy spectrum.  As the  magnetic field rotates the zero-magnetic field states move up or down relative to $E$ leading to jumps of $\Delta G_c=1$.  In the case of the hexagon, two facets can be parallel to the applied $B$-field.  This leads to two sets of such zero-magnetic field states which results in $\Delta G_c=2$, as is observed in Fig.\ \ref{fig:GTSH}c). 

As can be seen in Fig. \ref{fig:T_TSH}a (triangle) steps of $\Delta T=2$ are possible for certain energies (see $E=1.33$ for the clean wire).  But if the $G_c(\phi)$ can be measured for different energies, then it should possible to establish whether $\Delta T=2$ is predominant (hexagon) or $\Delta T=1$ is predominant (triangle).

\section{Conclusions}
\label{sec:conc}
We have studied surface electrons in nanowires made of a TI material, in the presence of an external magnetic field perpendicular to the longitudinal axis of the nanowire. The cross section of the nanowire is considered polygonal, as in most of the experimental realizations.  In the present work we have discussed the triangular, square, and hexagonal cases.  The component of the magnetic field normal to the electrons' trajectories depends on the facet of the prism, and consequently peculiar energy spectra are obtained, combining local Landau levels with free motion.  We have shown that the transmission function depends on the cross sectional geometry. And, naturally, for each geometry the conductance is anisotropic, i.e., it changes with the angular direction of the magnetic field.

These results can be useful to characterize the real TI nanowires fabricated in the labs \cite{Munning2021Feb,roessler23:2846}.  The presence of the conduction electrons on the surface and the polygonal cross section (often hexagonal) should lead to conductance features that we described, which should be robust in the presence of a modest amount of disorder. The sample quality can therefore be tested, for example by rotating the sample or the magnetic field (in a vector magnetic system), and comparing the conductance with the external geometry of the nanowire.

\begin{acknowledgments}
This research was supported by the Icelandic Research Fund, Grant 195943, the Knut and Alice Wallenberg Foundation (KAW) via the project Dynamic Quantum Matter (2019.0068), and from the Swedish Research Council (VR) through Grants No. 2019-04736 and No. 2020-00214.
\end{acknowledgments}

\appendix
\section{Facet orientation and ladder operators}
\label{app:Ladder}
We start with the linearized surface Hamiltonian in Eq.\ (\ref{eq:HsurfLinear}).  Intoducing the $k_x$-dependent center coordinate
\begin{equation}
    s_\pm(k_x) \equiv s_\pm+\frac{eA_x(s_\pm)+\hbar k_x}{eB^{\pm}},
\end{equation}
Eq.\ (\ref{eq:HsurfLinear}) can be written as
\begin{eqnarray}
H^{[\pm]}_\mathrm{surf}&=&v_F\left \{ \sigma_x\left [- eB_\pm
(s-s_\pm(k_x))
\right ]  +\sigma_y p_s\right \},
 \label{eq:HsurfLinear_App}
\end{eqnarray}
where $B^\pm=B \cos(\Phi_\pm)$.
Rewriting  Eq.\ (\ref{eq:HsurfLinear_App}) in terms of  $\sigma_\pm=\sigma_x \pm i \sigma_y$  results in
\begin{eqnarray}
H^{[\pm]}_\mathrm{surf}=v_F &&\left \{ \frac{\sigma_+}{2}
\left [- eB_\pm
(s-s_\pm(k_x))-ip_s
\right ]  \right . \nonumber  \\
& & \left . +\frac{\sigma_-}{2} 
\left [- eB_\pm
(s-s_\pm(k_x))+ip_s
\right ]
\right \}.
 \label{eq:Hsurf_sigmapm}
\end{eqnarray}
The sign of $\cos(\Phi_\pm)$ will determine the how the ladder operators are defined in terms of $s$ and $p_s$, as we show below.  For concreteness we now assume $\cos(\Phi_+) >0$.  
The requirement $[a,a^\dagger]=1$, fixes the definition of the ladder operator on that facet, i.e.\ 
\begin{equation}
a_{[+]}=\frac{\ell_c}{\sqrt{2} \hbar \sqrt{|\cos (\Phi_+)|}} \left [ -\frac{\hbar}{\ell_c^2}\cos( \Phi_+ )(s-s_+)+ip_s
\right ].
\label{eq:a_plus}
\end{equation}
In terms of these ladder operators the surface Hamiltonian of the facet is
\begin{equation}
H^{[+]}_\mathrm{surf}=\frac{\hbar v_F \sqrt{|\cos ( \Phi_+)|}}{\sqrt{2}\ell_c} \left (
 a_{[+]} \sigma_+ + a_{[+]}^\dagger \sigma_- \right ).
\label{eq:H_plus_App}
\end{equation}
The zero energy state of the above Hamiltonian is $|0\rangle |\!\!\downarrow \rangle$ since $ a_{[+]} \sigma_+ |0\rangle |\!\!\downarrow \rangle=0$, and $a_{[+]}^\dagger  \sigma_-|0 \rangle | \!\!\downarrow \rangle  =0$.

Repeating the same proceedure for the other facet results in the ladder operator
\begin{equation}
a_{[-]}=\frac{\ell_c}{\sqrt{2} \hbar \sqrt{|\cos( \Phi_-)}|} \left [ \frac{\hbar}{\ell_c^2}\cos (\Phi_-) (s-s_+) - ip_s
\right ],
\label{eq:a_minus}
\end{equation}
and the corresponding surface Hamiltonian of the facet
\begin{equation}
H^{[-]}_\mathrm{surf}=\frac{\hbar v_F \sqrt{|\cos ( \Phi_-)|}}{\sqrt{2}\ell_c} \left (
 a_{[-]}^\dagger \sigma_+ + a_{[-]} \sigma_- \right ).
\label{eq:H_minus_App}
\end{equation}
Again, the zero energy state of the above Hamiltonian is $|0\rangle |\!\!\uparrow \rangle$, since $ a_{[-]}^\dagger \sigma_+ |0\rangle |\!\!\uparrow \rangle=0$, and $a_{[-]}  \sigma_-|0 \rangle | \!\!\uparrow \rangle  =0$.

\section{Discretization with quadratic term}
\label{app:Discrete}
Starting from Eq.\ (\ref{eq:Hsurf_FD}), it can be written as
\begin{equation}
    H_\mathrm{surf}\psi(x,s) = H_0\psi(x,s) + V\psi(x+a,s) + V^\dagger\psi(x-a,s),
\end{equation}
where $H_0 = v_F\left(e\sigma_x\hat{A}_x(s) - i\hbar\sigma_y\partial_s\right)$ acts as an on-site potential of the lattice points and $V = - \frac{i\hbar v_F}{2a}\sigma_x$ acts as a nearest neighbour coupling term. 

Just as for the continuous model, the discrete model admits a simple analytical expression for its energy spectrum when the magnetic field is absent. Applying a Bloch ansatz $\psi(x,s)=e^{ik_x x}\phi_{k_x}(s)$ results in 
\begin{eqnarray}
H_\mathrm{surf}\psi(x,s) &=& \left(\hat{H}_0 + \hat{V}e^{ik_xa} + \hat{V}^\dagger e^{-ik_xa}\right) e^{ik_x x}\phi_{k_x}(s) \nonumber \\ 
&=& \hbar v_F\left(\frac{\sin(k_xa)}{a}\sigma_x + k_s\sigma_y\right)e^{ik_x x}\phi_{k_x}(s), \nonumber
\end{eqnarray}
which leads to eigenvalues 
\begin{equation}
E_\pm = \pm\hbar v_F\sqrt{\frac{\sin^2(k_xa)}{a^2} + k_s^2},
\end{equation}
which will converge to the spectrum of the continuous model in the limit $a\rightarrow 0$.   The anti-periodic boundary conditions is inherited by $\phi_{k_x}$, i.e., $\phi_{k_x}(s)=-\phi_{k_x}(s+P)$, which results in $k_s=\frac{2\pi}{P}(n+\frac{1}{2})$ with integer $n$.

There is however a serious issue with the discrete spectrum solution which is that it will close around the edges of the Brillouin zone ($k_x = \pm\pi/a$), effectively doubling the number of modes. This is the well known Fermion doubling problem. As it is an unwanted consequence of projecting our model onto a lattice, we remedy the problem by adding a Wilson mass term [Eq.\ (\ref{eq:H_wilson})] to our surface model.
The finite difference version of the quadratic term yields
\begin{eqnarray}
\partial_x^2\psi(x,s) \approx \frac{\left [ \psi(x+a,s) - 2\psi(x,s) + \psi(x-a,s)\right ] }{a^2},  \nonumber \\
\end{eqnarray}
such that the on-site potential and nearest neighour terms become 
\begin{eqnarray}
H_0 &=& v_F\left [ e\sigma_x\hat{A}_x(s) - i\hbar\sigma_y\partial_s + \frac{\beta}{\hbar}\sigma_z\left(\frac{2\hbar^2}{a^2} + e^2\hat{A}_x^2(s)\right) \right ], \nonumber \\
& &  \\
V &=& v_F\left [-\frac{i\hbar}{2a}\sigma_x + \frac{\beta}{\hbar}\sigma_z\left(-\frac{\hbar^2}{a^2} - \frac{i\hbar e\hat{A}_x(s)}{a}\right)\right].
\end{eqnarray}
For $B=0$, the plane wave solution now results in 
\begin{eqnarray}
H_\mathrm{surf}\psi(x,s) = \hbar v_F& &\biggl  [ \frac{\sin(k_xa)}{a}\sigma_x + 4\beta \frac{\sin^2(k_xa/2)}{a^2}\sigma_z  \nonumber   \\
& &  + k_s^2\sigma_y \biggr ] e^{ik_x x}\phi_{k_x}(s),
\end{eqnarray}
and corresponding eigenvalues 
\begin{equation}
E_\pm = \pm\hbar v_F\sqrt{\frac{\sin^2(k_xa)}{a^2} + 16\beta^2\frac{\sin^4(k_xa/2) }{a^4}+ k_s^2}.
\end{equation}
Taylor expanding the sinusoidal terms involving $k_x$ gives
\begin{equation}
\frac{\sin^2(k_xa)}{a^2} + 16\beta^2\frac{\sin^4(k_xa/2)}{a^4} \approx k_x^2 + \left(\beta^2 - \frac{a^2}{3}\right)k_x^4,
\end{equation}
which implies that the linear behaviour of the energy spectrum is best maintained by setting $\beta = \frac{a}{\sqrt{3}}$ 
 \cite{erlingsson18:1156}. In addition to maintaining the structure of the Dirac cone around the origin, the Wilson mass term succeeds in "lifting away" the spurious Dirac cone at the edges of the Brillouin zone. We can see this by considering the lowest energy modes at the edges where we have $E_\pm(k_x = \pm\pi/a, k_s = 0) = \pm\hbar v_F\frac{4}{a\sqrt{3}}$ such that the Dirac cone can always be lifted further away by decreasing the lattice constant $a$. Note also that in the limit of $a\rightarrow 0$ that the energy spectrum converges to the one of the continuum model.

 Having $H_0$ and $V$, along with an appropriate impurity potential $V_\mathrm{imp}$, defined they can then be written as matrices, indicated by $[\cdot]$, in a truncated basis of plane waves along the perimeter, and the spin basis. 
 The Green's function lattice equation then becomes
 \begin{eqnarray}
 (E[\mathbb{I}]-[H])G_{i,j}-[V]G_{i+1,j}-[V^\dagger]G_{i-1,j}=[\mathbb{I}]\delta_{i,j}, \nonumber \\
\end{eqnarray}
where $[H]=[H_0]+[V_\mathrm{imp}]$ and $i,j$ are lattice positions along the wire.
The above equation can then be implemented using standard recursive GF method \cite{ferry97:book,datta95:book,erlingsson17:036804}.


\end{document}